# Self-Dual Non-Abelian Vector Multiplet in Three Dimensions

Hitoshi NISHINO[1] and Subhash RAJPOOT[2]

*Department of Physics & Astronomy*
*California State University*
*1250 Bellflower Boulevard*
*Long Beach, CA 90840*

## Abstract

We present an $N=1$ supersymmetric non-Abelian compensator formulation for a vector multiplet in three-dimensions. Our total field content is the off-shell vector multiplet $(A_\mu{}^I, \lambda^I)$ with the off-shell scalar multiplet $(\varphi^I, \chi^I; F^I)$ both in the adjoint representation of an arbitrary non-Abelian gauge group. This system is reduced to a supersymmetric $\sigma$-model on a group manifold, in the zero-coupling limit. Based on this result, we formulate a 'self-dual' non-Abelian vector multiplet in three-dimensions. By an appropriate identification of parameters, the mass of the self-dual vector multiplet is quantized. Additionally, we also show that the self-dual non-Abelian vector multiplet can be coupled to supersymmetric Dirac-Born-Infeld action. These results are further reformulated in superspace to get a clear overall picture.

[1] E-Mail: hnishino@csulb.edu
[2] E-Mail: rajpoot@csulb.edu

## 1. Introduction

For theories with gauge invariance, so-called compensator fields or Stueckelberg fields [1] play interesting roles, such as giving masses to vector fields without Higgs fields [2]. The supersymmetrization of Stueckelberg formalism in four dimensions (4D) has been preformed already in 1970's [3], and also recently [4] both for Abelian gauge groups. Supersymmetric Abelian Stueckelberg formulations have been considered also for phenomenological applications [5]. However, these supersymmetric compensator formulations have been only for Abelian gauge groups.

As for the non-Abelian generalization of supersymmetric compensators, there seems to be certain obstruction at least in 4D. The origin of such an obstruction seems to be due to the limitation of available multiplets in 4D. For a desirable compensator, we have to have a spin 0 field in the adjoint representation. The trouble is that such spin 0 fields can be found only in a chiral multiplet $(A, B, \psi)$ or a tensor multiplet $(B_{\mu\nu}, \chi, \varphi)$. The latter is problematic, because the tensor should be also in the adjoint representation, *i.e.*, the problematic non-Abelian tensor [6]. On the other hand, in a chiral multiplet, both the scalar $A$ and the pseudoscalar $B$ are in the adjoint representation. However, in order for the $A$-field to be exponentiated as a compensator field, it is very difficult to separate the $B$-field not to interfere with the compensator, maintaining supersymmetry. We can try to complexify the gauge transformation parameter, but the price to be paid is the complexification of vector field which costs the doubling of the vector multiplet (VM). This problem is more transparent in superspace: If we try to exponentiate the chiral superfield as a compensator superfield, then the $B$-field will be inevitably involved non-linearly, and it is difficult to separate the $B$-field from the $A$-field. Considering these points, it seems almost impossible to formulate supersymmetric compensators for non-Abelian gauge groups.

There is, however, one way to circumvent this problem, by changing the space-time dimensions. Instead of dealing with problematic 4D case, we can work in 3D for the $N=1$ supersymmetrization of compensators for non-Abelian gauge groups. This is possible because a scalar multiplet in 3D has only a single scalar. In this paper, we will introduce the $N=1$ off-shell VM $(A_\mu{}^I, \lambda^I)$ in the adjoint representation of an arbitrary gauge group $G$, and an off-shell scalar multiplet (SM) $(\varphi^I, \chi^I; F^I)$ also in the adjoint representation. We will adopt the off-shell SM with the auxiliary field $F^I$. The VM stays within the Wess-Zumino gauge with no auxiliary fields needed. Interestingly, we will see that the compensator field strength $P_\mu \equiv (D_\mu e^\varphi)e^{-\varphi}$ plays a crucial role also for the supersymmetric consistency of the total system. As one of the most important applications, we use this result to formulate a non-Abelian 'self-dual' VM in 3D which had been considered to be extremely difficult ever since the original work in 1980's [7], except for sophisticated theory such as $N = 4$ supergravity in 7D [8].

We also see that our system has a close relationship with $\sigma$-models for gauge group manifolds, when the minimal coupling is switched off: $m \to 0$, justifying various coefficients in our lagrangians. Subsequently, we also show that the self-dual VM can be further coupled to supersymmetric Dirac-Born-Infeld (DBI) action [9][10][11]. By an appropriate identification of mass parameters, we will see that the mass of the self-dual VM can be quantized as the result of Chern-Simons quantization. Subsequently, we also couple the self-dual massive VM to supersymmetric DBI action. Finally, we give the reformulation of these component results in superspace that might provide a clearer picture for the whole subject.

## 2. Compensators and Gauge Covariance

We first establish the right way to describe the compensators for an arbitrary non-Abelian gauge group $G$. Let $\varphi \equiv \varphi^I T^I$ be the set of Lie-ring-valued scalar fields in the adjoint representation with the anti-hermitian generators $T^I$, where $I = 1, 2, \cdots, \dim G \equiv g$. The anti-hermitian generators satisfy the commutator

$$[T^I, T^J] = f^{IJK} T^K \quad , \tag{2.1}$$

where $f^{IJK}$ is the structure constants. Let $A_\mu{}^I$ be the gauge fields for the gauge group $G$, whose field strength is defined by

$$F_{\mu\nu} \equiv \partial_\mu A_\nu - \partial_\nu A_\mu + m[A_\mu, A_\nu] \quad . \tag{2.2}$$

As in this expression, we sometimes omit adjoint indices in order to make the expressions simpler. The $m$'s is the gauge-coupling constant with the dimension of mass.[3] The finite gauge transformations for these fields will be[4]

$$(e^\varphi)' = e^{-\Lambda} e^\varphi \quad , \quad (e^{-\varphi})' = e^{-\varphi} e^\Lambda \quad , \tag{2.3a}$$

$$A_\mu{}' = m^{-1} e^{-\Lambda} \partial_\mu e^\Lambda + e^{-\Lambda} A_\mu e^\Lambda \quad , \tag{2.3b}$$

$$F_{\mu\nu}{}' = e^{-\Lambda} F_{\mu\nu} e^\Lambda \quad , \tag{2.3c}$$

---

[3] In this paper we assign the physical engineering dimension 0 (or 1/2) to the bosons (or fermions). Accordingly, the gauge coupling constant $m$ has the dimension of mass.

[4] These transformation rules have been known in the past, *e.g.*, [2].

where $\Lambda \equiv \Lambda^I(x) T^I$ are $x$-dependent finite local gauge transformation parameters. Needless to say, all the terms in (2.3) are Lie-ring valued.

We can now define the covariant derivative acting on $e^\varphi$ by [2]

$$D_\mu e^\varphi \equiv \partial_\mu e^\varphi + m A_\mu e^\varphi ~, \tag{2.4}$$

transforming 'covariantly' under (2.3):

$$(D_\mu e^\varphi)' = e^{-\Lambda}(D_\mu e^\varphi) ~. \tag{2.5}$$

Relevantly, we can define the covariant field strength for $\varphi$ by

$$P_\mu \equiv (D_\mu e^\varphi) e^{-\varphi} ~, \tag{2.6}$$

transforming as

$$P_\mu' = e^{-\Lambda} P_\mu e^\Lambda ~. \tag{2.7}$$

Accordingly, it is convenient to have the arbitrary infinitesimal variation

$$\delta P_\mu = [D_\mu - P_\mu, (\delta e^\varphi) e^{-\varphi}] + m \delta A_\mu ~. \tag{2.8}$$

This implies that the product $P_\mu{}^I P^{\mu I}$ is the most appropriate choice for a gauge-covariant kinetic term for the $\varphi$-field. Relevantly, an important identity is

$$D_{[\mu} P_{\nu]} = +\tfrac{1}{2} m F_{\mu\nu} + \tfrac{1}{2}[P_\mu, P_\nu] ~. \tag{2.9}$$

We can now understand the role of the compensator scalars by the 'toy' lagrangian[5]

$$\mathcal{L}_{\text{toy}}(x) = -\tfrac{1}{4}(F_{\mu\nu}{}^I)^2 - \tfrac{1}{2}(P_\mu{}^I)^2 ~. \tag{2.10}$$

We now redefine the gauge field by

$$\widetilde{A}_\mu \equiv e^{-\varphi} A_\mu e^\varphi + m^{-1} e^{-\varphi}(\partial_\mu e^\varphi) = m^{-1} e^{-\varphi} P_\mu e^\varphi ~, \tag{2.11}$$

so that the new field $\widetilde{A}_\mu$ and its field strength do *not* transform [1][2]

$$\widetilde{A}_\mu' = \widetilde{A}_\mu ~, \quad \widetilde{F}_{\mu\nu}' = \widetilde{F}_{\mu\nu} ~. \tag{2.12}$$

Relevantly, the inverse relationships are

$$P_\mu = m e^\varphi \widetilde{A}_\mu e^{-\varphi} ~, \quad F_{\mu\nu} = e^\varphi \widetilde{F}_{\mu\nu} e^{-\varphi} ~. \tag{2.13}$$

---

[5] Our metric in this paper is $(\eta_{\mu\nu}) = \text{diag.}(-,+,+)$.

Note that the exponential factors $e^{\pm\varphi}$ are entirely absent from the lagrangian:

$$\mathcal{L}_{\text{toy}}(x) = -\tfrac{1}{4}(\widetilde{F}_{\mu\nu}{}^I)^2 - \tfrac{1}{2}m^2(\widetilde{A}_\mu{}^I)^2 \ . \tag{2.14}$$

In other words, the original kinetic term for $\varphi$ is now reduced to the mass term of $\widetilde{A}_\mu$ [1][2]. Now the original gauge invariance of the action is no longer manifest, because at the moment the scalars $\varphi$ are absorbed, the gauge degree of freedom is fixed. Even though these features of compensators have been known in the past [1][2], we now consider their supersymmetrization.

## 3.  $N=1$  Supersymmetric Compensator SM in 3D

With the preliminaries above, we are ready for presenting the invariant action for the VM + SM, where the latter is the compensator SM. Our total action $I_1 \equiv I_{\text{VM}} + I_{\text{SM}}$ has the corresponding lagrangians

$$\mathcal{L}_{\text{VM}}(x) = -\tfrac{1}{4}(F_{\mu\nu}{}^I)^2 - \tfrac{1}{2}(\overline{\lambda}^I \slashed{D} \lambda^I) \ , \tag{3.1a}$$

$$\mathcal{L}_{\text{SM}}(x) = -\tfrac{1}{2}(P_\mu{}^I)^2 - \tfrac{1}{2}(\overline{\chi}^I \slashed{\mathcal{D}} \chi^I) + \tfrac{1}{2}(F^I)^2 - m(\overline{\lambda}^I \chi^I) + \tfrac{1}{48} h^{IJ,KL}(\overline{\chi}^I \chi^K)(\overline{\chi}^J \chi^L) \ , \tag{3.1b}$$

where $h^{IJ,KL} \equiv f^{IJM} f^{MKL}$. The covariant derivatives $D_\mu$ and $\mathcal{D}_\mu$ are defined by

$$D_\mu \lambda^I \equiv \partial_\mu \lambda^I + m f^{IJK} A_\mu{}^J \lambda^K \ , \tag{3.2a}$$

$$\mathcal{D}_\mu \chi^I \equiv D_\mu \chi^I - \tfrac{1}{2} m f^{IJK} P_\mu{}^J \chi^K \ . \tag{3.2b}$$

Note the peculiar coefficient '$-1/2$' in the last $P\chi$-term.

Each of the actions $I_{\text{VM}}$ and $I_{\text{SM}}$ is separately invariant under off-shell supersymmetry

$$\delta_Q A_\mu{}^I = +(\overline{\epsilon} \gamma_\mu \lambda^I) \ , \tag{3.3a}$$

$$\delta_Q \lambda^I = -\tfrac{1}{2}(\gamma^{\mu\nu}\epsilon) F_{\mu\nu} \ , \tag{3.3b}$$

$$\delta_Q e^\varphi = +(\overline{\epsilon}\chi) e^\varphi \ , \tag{3.3c}$$

$$\delta_Q \chi^I = +(\gamma^\mu \epsilon) \left[ P_\mu{}^I - \tfrac{1}{4} f^{IJK}(\overline{\chi}^J \gamma_\mu \chi^K) \right] + \epsilon F^I \ , \tag{3.3d}$$

$$\delta_Q F^I = +(\overline{\epsilon}\slashed{\mathcal{D}}\chi^I) + m(\overline{\epsilon}\lambda^I) - \tfrac{1}{12} h^{IJ,KL}(\overline{\epsilon}\chi^K)(\overline{\chi}^J \chi^L) \ . \tag{3.3e}$$

In (3.3c), both sides are Lie-ring valued. Note that (3.3a) and (3.3b) are within the Wess-Zumino gauge with no auxiliary fields.

The component field equations are obtained from the total action $I_1$ as[6]

$$\frac{\delta I_1}{\delta \lambda^I} = -\slashed{D}\lambda^I - m\chi^I \doteq 0 ~, \tag{3.4a}$$

$$\frac{\delta I_1}{\delta \chi^I} = -\slashed{D}\chi^I - m\lambda^I + \tfrac{1}{12} h^{IJ,KL} \chi^K (\overline{\chi}^J \chi^L) \doteq 0 ~, \tag{3.4b}$$

$$\frac{\delta I_1}{\delta A_\mu{}^I} = -D_\nu F^{\mu\nu\,I} - mP_\mu{}^I + \tfrac{1}{2} m f^{IJK} (\overline{\lambda}^J \gamma_\mu \lambda^K) + \tfrac{1}{4} m f^{IJK} (\overline{\chi}^J \gamma^\mu \chi^K) \doteq 0 ~, \tag{3.4c}$$

$$e^\varphi \frac{\delta I_1}{\delta e^\varphi} = \left[ +D_\mu P^{\mu\,I} + \tfrac{1}{2} f^{IJK} (\overline{\chi}^J \slashed{D}\chi^K) - \tfrac{1}{8} h^{IJ,KL} (\overline{\chi}^K \gamma_\mu \chi^L) P_\mu{}^J \right] T^I \doteq 0 ~, \tag{3.4d}$$

$$\frac{\delta I_1}{\delta F^I} = +F^I \doteq 0 ~. \tag{3.4e}$$

We can also confirm the supercovariance of these component field equations under (3.3).

We now consider the absorption of the compensator $\varphi$ into the vector field, so that the latter becomes explicitly massive. The prescription is basically (2.11), so that the total lagrangian $\mathcal{L}_1$ becomes

$$\mathcal{L}_1(x) = -\tfrac{1}{4}(\widetilde{F}_{\mu\nu}{}^I)^2 - \tfrac{1}{2}(\overline{\widetilde{\lambda}}^I \slashed{D} \widetilde{\lambda}^I) - \tfrac{1}{2} m^2 (\widetilde{A}_\mu{}^I)^2 - \tfrac{1}{2}(\overline{\widetilde{\chi}}^I \slashed{D} \widetilde{\chi}^I) + \tfrac{1}{2}(\widetilde{F}^I)^2$$
$$- m(\overline{\widetilde{\lambda}}^I \widetilde{\chi}^I) + \tfrac{1}{48} h^{IJ,KL} (\overline{\widetilde{\chi}}^I \widetilde{\chi}^K)(\overline{\widetilde{\chi}}^J \widetilde{\chi}^L) ~, \tag{3.5}$$

and the corresponding supersymmetry transformation rule is

$$\delta_Q \widetilde{A}_\mu = +(\overline{\epsilon}\gamma_\mu \widetilde{\lambda}) + \widetilde{D}_\mu(\overline{\epsilon}\widetilde{\chi}) ~, \tag{3.6a}$$

$$\delta_Q \widetilde{\lambda}^I = -\tfrac{1}{2}(\gamma^{\mu\nu}\epsilon)\widetilde{F}_{\mu\nu} - [(\overline{\epsilon}\widetilde{\chi}), \widetilde{\lambda}] ~, \tag{3.6b}$$

$$\delta_Q \widetilde{\chi}^I = +(\gamma^\mu \epsilon)\left[ m\widetilde{A}_\mu{}^I - \tfrac{1}{4} f^{IJK}(\overline{\widetilde{\chi}}^J \gamma_\mu \widetilde{\chi}^K) \right] + \epsilon \widetilde{F}^I - [(\overline{\epsilon}\widetilde{\chi}), \widetilde{\chi}] ~, \tag{3.6c}$$

$$\delta_Q \widetilde{F}^I = +(\overline{\epsilon}\slashed{D}\widetilde{\chi}^I) + m(\overline{\epsilon}\widetilde{\lambda}^I) - \tfrac{1}{12} h^{IJ,KL} (\overline{\epsilon}\widetilde{\chi}^K)(\overline{\widetilde{\chi}}^J \widetilde{\chi}^L) - [(\overline{\epsilon}\widetilde{\chi}), \widetilde{F}] ~, \tag{3.6d}$$

where

$$(\widetilde{\lambda}, \widetilde{\chi}, \widetilde{F}) \equiv e^{-\varphi}(\lambda, \chi, F)e^\varphi ~, \tag{3.7a}$$

$$\widetilde{D}_\mu \widetilde{\lambda} \equiv \partial_\mu \widetilde{\lambda} + m[\widetilde{A}_\mu, \widetilde{\lambda}] ~, \quad \widetilde{D}_\mu \widetilde{\chi} \equiv \partial_\mu \widetilde{\chi} + m[\widetilde{A}_\mu, \widetilde{\chi}] ~, \tag{3.7b}$$

$$\widetilde{\mathcal{D}}_\mu \widetilde{\chi} \equiv \partial_\mu \widetilde{\chi} + \tfrac{1}{2} m[\widetilde{A}_\mu, \widetilde{\chi}] ~. \tag{3.7c}$$

Since the $\varphi$'s has been totally absorbed into $\widetilde{A}_\mu$, no $\varphi$-transformation is needed any longer. All the commutators in (3.6b) through (3.6d) and the covariant derivative in (3.6a) can be interpreted as compensating gauge transformation from the *untilded* fields into *tilded* fields.

---

[6] We use the special symbol $\doteq$ for a field equation in this paper.

We thus see that the peculiar coefficients for terms with $\widetilde{A}_\mu$ have been fixed, which could not have been fixed so easily without the gauge invariance with compensator fields.

## 4. Relationship with Group Manifold $\sigma$-Model for $m \to 0$

Note the important limit of $m \to 0$, when the minimal coupling is switched off. Even in this case, there are non-trivial interactions within the SM, because the non-trivial kinetic terms of $\varphi$ and $\chi$ remain. The $\varphi$-kinetic term becomes nothing but the usual $\sigma$-model kinetic term for the group manifold, with the $N = 1$ supersymmetric partner kinetic term of $\chi$ with the couplings through $P_\mu$.

To be more specific, the equation relating our original notation and the conventional $\sigma$-model notation for the $m = 0$ case is

$$P_\mu|_{m=0} = (\partial_\mu e^\varphi) e^{-\varphi} \equiv -(\partial_\mu \phi^\alpha) e_\alpha{}^I T^I \ , \tag{4.1}$$

where $\phi^\alpha$ is the $\sigma$-model coordinates with the curved-coordinate index $\alpha = 1, 2, \cdots, g \equiv \dim G$. Note the negative sign in the r.h.s. Relevantly, other important key relationships are such as[7]

$$g_{\alpha\beta} = e_\alpha{}^I e_\beta{}^I \ , \tag{4.2a}$$

$$(\delta e^\varphi) e^{-\varphi} = -(\delta \phi^\alpha) e_\alpha{}^I T^I \ , \tag{4.2b}$$

$$\omega_\alpha{}^{IJ} = -\tfrac{1}{2} f_\alpha{}^{IJ} \equiv -\tfrac{1}{2} f^{KIJ} e_\alpha{}^K \ , \tag{4.2c}$$

$$C_{\alpha\beta}{}^I \equiv \partial_\alpha e_\beta{}^I - \partial_\beta e_\alpha{}^I = -f_{\alpha\beta}{}^I \equiv -f^{JKI} e_\alpha{}^J e_\beta{}^K \ . \tag{4.2d}$$

$$R_{\alpha\beta}{}^{IJ} \equiv (\partial_\alpha \omega_\beta{}^{IJ} + \omega_\alpha{}^{IK} \omega_\beta{}^{KJ}) - (\alpha \leftrightarrow \beta) = \tfrac{1}{4} h_{\alpha\beta}{}^{IJ} \equiv \tfrac{1}{4} e_\alpha{}^K e_\beta{}^L h^{KL,IJ} \ , \tag{4.2e}$$

$$T_{\alpha\beta}{}^I \equiv \partial_\alpha e_\beta{}^I - \partial_\beta e_\alpha{}^I + \omega_\alpha{}^{IJ} e_\beta{}^J - \omega_\beta{}^{IJ} e_\alpha{}^J = 0 \ , \tag{4.2f}$$

where $g_{\alpha\beta}$ is the metric tensor on the group manifold, $e_\alpha{}^I$ is its vielbein, $\omega_\alpha{}^{IJ}$ is the Lorentz connection, $C_{\alpha\beta}{}^I$ is the anholonomy coefficient, $R_{\alpha\beta}{}^{IJ}$ is the Riemann curvature tensor, and $T_{\alpha\beta}{}^I$ is the torsion tensor that vanishes. Some of these relationships have been known in the context of Kaluza-Klein theories [12]. Needless to say, the vielbein $e_\alpha{}^I$ satisfies the 'vielbein postulate':

$$D_\alpha e_\beta{}^I \equiv \partial_\alpha e_\beta{}^I + \omega_\alpha{}^{IJ} e_\beta{}^J - \left\{{}^{\gamma}_{\alpha\ \beta}\right\} e_\gamma{}^I \equiv 0 \ . \tag{4.3}$$

---

[7] Since the gauge group we are dealing with has always positive definite metric $\delta^{IJ}$, we always use the superscripts for the local indices $I, J, \cdots$. As for curved indices $\alpha, \beta, \cdots$, we distinguish their upper/lower cases, because of the involvement of $g_{\alpha\beta}(\phi)$ or its inverse.

After simple manipulations based on these relationships, we can easily show that the lagrangian $\mathcal{L}_{\rm SM}(x)$ in (3.1b) with $m \to 0$ becomes

$$\mathcal{L}_{\rm SM}(x)|_{m \to 0} = -\tfrac{1}{2} g_{\alpha\beta} (\partial_\mu \phi^\alpha)(\partial^\mu \phi^\beta) - \tfrac{1}{2} g_{\alpha\beta} (\overline{\chi}^\alpha \slashed{D} \chi^\beta) + \tfrac{1}{2} g_{\alpha\beta} F^\alpha F^\beta$$
$$+ \tfrac{1}{12} R_{\alpha\beta\gamma\delta} (\overline{\chi}^\alpha \chi^\gamma)(\overline{\chi}^\beta \chi^\delta) \ , \tag{4.4}$$

where $\chi^\alpha \equiv \chi^I e^{I\alpha}$, $F^\alpha \equiv F^I e^{I\alpha}$, and the covariant derivative $D_\mu$ is with the Christoffel symbol:

$$D_\mu \chi^\alpha \equiv \partial_\mu \chi^\alpha + (\partial_\mu \phi^\beta) \left\{ {}^{\ \alpha}_{\beta\ \gamma} \right\} \chi^\gamma \ . \tag{4.5}$$

Eq. (4.4) is nothing but the $N=1$ supersymmetric $\sigma$-model lagrangian based on the group manifold as a torsionless Riemannian manifold. This is another by-product of our supersymmetric compensator formulation in 3D.

The relationship with the $\sigma$-model above provides an additional confirmation of various coefficients in our lagrangian, in particular, the special $\slashed{P}\chi$-term in $\mathcal{D}_\mu \chi$. It is this coefficient that explains the minimal coupling of the Lorentz connection in the $\chi$-kinetic term via (4.2c):

$$\mathcal{D}_\mu \chi^I |_{m=0} = \partial_\mu \chi^I + (\partial_\mu \phi^\alpha) \omega_\alpha{}^{IJ} \chi^J \ . \tag{4.6}$$

## 5. $N=1$ Supersymmetric Non-Abelian Self-Dual VM in 3D

The new concept of 'self-duality' in odd dimensions had been introduced in [7] for an Abelian gauge field. However, the generalization of this system to non-Abelian system has been supposed to be extremely difficult. One of the reasons seems to be the lack of gauge invariance of the total action, so that possible terms in the lagrangian increases, and we lose the control of the supersymmetrization of the system.

This problem can be now solved, because we have established the $N=1$ compensator SM. In other words, we can temporarily keep the gauge invariance of the total action, before we choose the special gauge in which the scalar fields are absorbed into the field redefinition of the non-Abelian gauge field, as we have seen in the toy lagrangian (2.14).

Our total action is now $I_2 \equiv \int d^3x \, \mathcal{L}_2(x)$ with $\mathcal{L}_2 \equiv \mathcal{L}_{\rm SCS} + \mathcal{L}_{\rm SM}$, where $\mathcal{L}_{\rm SM}$ is the same as (3.1b), while $\mathcal{L}_{\rm SCS}$ is the supersymmetric Chern-Simons lagrangian

$$\mathcal{L}_{\rm SCS}(x) = +\tfrac{1}{4} \mu \, \epsilon^{\mu\nu\rho} (F_{\mu\nu}{}^I A_\rho{}^I - \tfrac{1}{3} m f^{IJK} A_\mu{}^I A_\nu{}^J A_\rho{}^K) - \tfrac{1}{2} \mu (\overline{\lambda}^I \lambda^I) \ . \tag{5.1}$$

The mass parameter $\mu$ can be arbitrary and independent of $m$. Since we are dealing with off-shell supersymmetry (3.3), the new action $I_{\text{SCS}}$ is by itself invariant under (3.3) which is not modified.

Our field equations are now

$$\frac{\delta I_2}{\delta \lambda^I} = -\mu \lambda^I - m\chi^I \doteq 0 \ , \tag{5.2a}$$

$$\frac{\delta I_2}{\delta \chi^I} = -\slashed{D}\chi^I - \mu\lambda^I + \tfrac{1}{12} h^{IJ,KL}\chi^K(\overline{\chi}^J \chi^L) \doteq 0 \ , \tag{5.2b}$$

$$\frac{\delta I_2}{\delta A_\mu{}^I} = +\tfrac{1}{2}\mu\, \epsilon^{\mu\rho\sigma} F_{\rho\sigma}{}^I - m P_\mu{}^I + \tfrac{1}{4} m f^{IJK}(\overline{\chi}^J \gamma^\mu \chi^K) \doteq 0 \ , \tag{5.2c}$$

$$e^\varphi \frac{\delta I_2}{\delta e^\varphi} = \left[ +D_\mu P^{\mu I} + \tfrac{1}{2} f^{IJK}(\overline{\chi}^J \slashed{D}\chi^K) - \tfrac{1}{8} h^{IJ,KL}(\overline{\chi}^K \gamma_\mu \chi^L) P_\mu{}^J \right] T^I \doteq 0 \ , \tag{5.2d}$$

$$\frac{\delta I_2}{\delta F^I} = +F^I \doteq 0 \ . \tag{5.2e}$$

The most important equation is (5.2c), which yields the self-duality condition in 3D, after the scalar fields $\varphi$ are absorbed into the new field $\widetilde{A}_\mu$ in (2.11). This is because $P_\mu{}^I = m \widetilde{A}_\mu{}^I$ is proportional to a mass term for $\widetilde{A}_\mu$, yielding the non-Abelian version of the self-duality condition

$$\tfrac{1}{2}\mu\, \epsilon_\mu{}^{\rho\sigma} \widetilde{F}_{\rho\sigma}{}^I \doteq +m^2 \widetilde{A}_\mu{}^I - \tfrac{1}{4} m f^{IJK}(\overline{\widetilde{\chi}}^J \gamma_\mu \widetilde{\chi}^K) \ . \tag{5.3}$$

Since the $\mu$-coefficient is independent of $m$, we have an additional freedom for controlling the self-duality here. The fermionic bilinear term in our self-dual condition (5.3) is analogous to usual self-duality conditions in supersymmetric theories in even dimensions, such as self-dual tensor in 6D [13] or in other higher-dimensions [14].

The coefficient $\mu$ in (5.1) is to be quantized for non-Abelian gauge groups with non-trivial $\pi_3$-homotopy mappings [15]. These mappings are given by

$$\pi_3(G) = \begin{cases} \mathbb{Z} & (\text{for } G = A_i,\ B_i,\ C_i,\ D_i\ (i \geq 2,\ G \neq D_2),\ G_2,\ F_4,\ E_6,\ E_7,\ E_8) \ , \\ \mathbb{Z} \oplus \mathbb{Z} & (\text{for } G = SO(4)) \ , \\ 0 & (\text{for } G = U(1)) \ . \end{cases} \tag{5.4}$$

For $\forall G$ with $\pi_3(G) = \mathbb{Z}$, the coefficient $\mu$ is quantized as

$$\mu = \frac{n}{4\pi} \quad (n = 0, \pm 1, \pm 2, \cdots) \ . \tag{5.5}$$

Accordingly, this quantization condition is involved in the self-duality equation (5.3).

The quantization of the $\mu$-coefficient is purely non-Abelian feature of the system, which did not arise in the Abelian case in the past [7]. In other words, our system is the first one in 3D that has both self-duality and the quantization of the $\mu$-coefficient in (5.3). Note also that by the identification $\mu \equiv m$, the mass of the self-dual vector itself is quantized as $4\pi m = n \in \mathbb{Z}$.

## 6. Coupling to Supersymmetric DBI Action

Since our formulation has been in off-shell component language, it is not too difficult to introduce more general interactions, such as DBI action [9]. As a matter of fact, we can consider the supersymmetric DBI action [10] in 3D [11] at the first non-trivial quartic order:

$$\mathcal{L}_{\text{SDBI}}(x) = \alpha^2 \text{ Str} \Big[ + (\widehat{F}_\mu)^2(\widehat{F}_\nu)^2 - 2(\overline{\lambda}\slashed{D}\lambda)(\widehat{F}_\mu)^2 + \widehat{F}_\mu \overline{\lambda} \gamma^\mu \slashed{D} (\gamma^\nu \lambda \widehat{F}_\nu) \\ + (\overline{\lambda}\slashed{D}\lambda)^2 + \tfrac{1}{4}(\overline{\lambda}\lambda) D_\mu^2 (\overline{\lambda}\lambda) \Big] + \mathcal{O}(\phi^5) ~, \tag{6.1}$$

where $\mathcal{O}(\phi^5)$ is for terms at the quintic order in fields, and $\alpha$ is nonzero real constant with the dimension of (mass)$^{-1}$, while all the fields carry the generators. The $\widehat{F}_\mu \equiv (1/2)\epsilon_\mu{}^{\rho\sigma} F_{\rho\sigma}$ is the dual of $F_{\rho\sigma}$. The symbol 'Str' is for the totally symmetrized trace operation, i.e.,

$$\text{Str}\,(T^I T^J T^K T^L) = \text{ tr}\,[T^{(I} T^J T^K T^{L)}] \equiv C^{IJKL} ~, \tag{6.2}$$

so that the coefficient $C^{IJKL}$ is totally symmetric, whose explicit values depend on the group $G$ [16]. Since the VM has no auxiliary field even off-shell, we do not have any auxiliary-field dependent term in (6.1). Additionally, the supersymmetry transformation rules (3.3a) and (3.3b) do not change, even after $\mathcal{L}_{\text{SDBI}}$ is added.

The important conceptual point is that the vector field equation from the total action $I_3 \equiv I_{\text{SCS}} + I_{\text{SM}} + I_{\text{SDBI}}$ is now

$$\frac{\delta I_3}{\delta A_\mu{}^I} = + \tfrac{1}{2}\mu\,\epsilon^{\mu\nu\rho} F_{\nu\rho}{}^I - m P_\mu{}^I + \tfrac{1}{4} m f^{IJK}(\overline{\chi}^J \gamma^\mu \chi^K) \\ + \alpha^2 \epsilon^{\mu\rho\sigma} C^{IJKL} D_\rho \Big[ 4 \widehat{F}_\sigma{}^J \widehat{F}_\tau{}^K \widehat{F}^{\tau L} - 4(\overline{\lambda}^J \slashed{D} \lambda^K) \widehat{F}_\sigma{}^L \\ + 2 \overline{\lambda}^J \gamma_\sigma \slashed{D}(\gamma^\tau \lambda^K \widehat{F}_\tau{}^L) \Big] + \mathcal{O}(\phi^4) \doteq 0 ~, \tag{6.3}$$

containing the mass term for the vector interacting with the supersymmetric DBI-terms, after absorbing the compensator into $P_\mu{}^I \equiv m \widetilde{A}_\mu{}^I$. To our knowledge, such massive supersymmetric DBI interactions have never given explicitly in the past.

We have thus accomplished not only the generalization of $N=1$ self-dual VM in 3D to non-Abelian case, but also the coupling it to a DBI action. Even though the Abelian supersymmetric DBI action was introduced to self-dual VM [11], our result (6.3) covers the non-Abelian case for the first time after the discovery of self-dual VM in 3D [7].

## 7. Superspace Formulation

Once we have established the component formulation of the compensator SM, the next natural step is to reformulate these results in superspace [17][18]. In fact, this turns out to be not too difficult. We give here the main results needed for such superspace reformulation.

We start with the basic superfield relationships. The SM $(\varphi^I, \chi^I; F^I)$ is now embedded in the scalar superfield $\Phi \equiv \Phi^I T^I$ in the adjoint representation. The gauge supercovariant derivatives $(\nabla_A) \equiv (\nabla_a, \nabla_\alpha)$ are defined by[8])

$$\nabla_a \equiv \partial_a + mA_a \ , \quad \nabla_\alpha \equiv D_\alpha + mA_\alpha \ , \tag{7.1}$$

where all the superfields are Lie-ring valued, and $A_\alpha$ is the fundamental spinor superfield, corresponding to $\Gamma_\alpha$ in [17]. Accordingly, the superfield strength $P_A$ is now defined by

$$P_A \equiv (\nabla_A e^\Phi) e^{-\Phi} \ , \quad \nabla_A e^\Phi \equiv D_A e^\Phi + mA_A e^\Phi \ . \tag{7.2}$$

Under a finite local non-Abelian transformation, they transform as

$$(e^\Phi)' = e^{-\Lambda} e^\Phi \ , \quad (e^{-\Phi})' = e^{-\Phi} e^\Lambda \ , \tag{7.3a}$$

$$A_A' = e^{-\Lambda} D_A e^\Lambda + m e^{-\Lambda} A_A e^\Lambda \ , \tag{7.3b}$$

$$P_A' = e^{-\Lambda} P_A e^\Lambda \ , \tag{7.3c}$$

where $\Lambda \equiv \Lambda^I(z) T^I$ is now a scalar parameter superfield for a finite non-Abelian gauge transformation. Even though these are parallel to the component cases in earlier sections, superspace reformulation provides a clearer picture of the total project. The correspondence to the component fields is such as

$$\Phi| = \varphi \ , \quad P_\alpha| = -\chi_\alpha \ , \quad -\tfrac{1}{2} \nabla^\alpha P_\alpha| = F \ . \tag{7.4}$$

We skip the relationships for VM which have been already known [17].

---

[8]) For superspace coordinates $(z^A) \equiv (x^a, \theta^\alpha)$, we use the indices $A \equiv (a, \alpha)$, $B \equiv (b, \beta)$, $\cdots$, where $a, b, \cdots = 0, 1, 2$ (or $\alpha, \beta, \cdots = 1, 2$) are for the bosonic (or fermionic) coordinates. We also maintain the signature $(\eta_{ab}) = \text{diag.} (-, +, +)$, which generates a slight difference from [17]. We use the convention that all the Lie-ring valued quantities $X$ in terms of anti-hermitian generators: $X \equiv X^I T^I$ in contrast to the hermitian generators in [17]. Other conventions, such as $A_{\alpha\beta} \equiv (\gamma^c)_{\alpha\beta} A_c$ are the same as in [17].

We now give the superspace lagrangians for $I_{\rm SM}$, $I_{\rm VM}$ and $I_{\rm SCS}$, where the latter two have been well-known for some time [17], while the first one is our new result here:

$$I_{\rm SM} = \int d^3x\, d^2\theta\, \mathcal{L}_{\rm SM}(z) = \int d^3x\, d^2\theta \left(-\tfrac{1}{4} P^{\alpha I} P_\alpha{}^I\right) , \tag{7.5a}$$

$$I_{\rm VM} = \int d^3x\, d^2\theta\, \mathcal{L}_{\rm VM}(z) = \int d^3x\, d^2\theta\, c_1 \left(W^{\alpha I} W_\alpha{}^I\right) , \tag{7.5b}$$

$$I_{\rm SCS} = \int d^3x\, d^2\theta\, \mathcal{L}_{\rm SCS}(z) = \int d^3x\, d^2\theta\, c_2\, \mu\, A^{\alpha I} \left(W_\alpha{}^I - \tfrac{i}{6} m [A^\beta, A_{\alpha\beta}]^I\right) . \tag{7.5c}$$

The real constants $c_1$ and $c_2$ depends on the normalizations in the VM sector. Here, as usual, $W_\alpha{}^I$ is defined in terms of $A_\alpha$ [17] by[9]

$$W_\alpha = \tfrac{1}{2} D^\beta D_\alpha A_\beta + \tfrac{1}{2} m [A^\beta, D_\beta A_\alpha] + \tfrac{1}{6} m^2 [A^\beta, \{A_\beta, A_\alpha\}] , \tag{7.6}$$

The crucial relationship is

$$\nabla^\alpha \nabla_\alpha P_\beta | = -2(\slashed{\mathcal{D}}\chi)_\beta - 4m\lambda_\beta , \tag{7.7}$$

where the peculiar covariant derivative $\mathcal{D}_a \chi$ arises in the r.h.s. consistently with the component result (3.2b), with the special coefficient '$-1/2$' for the $P\chi$-term in $\mathcal{D}_a\chi$. Using these equations, we can re-obtain the component lagrangian $\mathcal{L}_{\rm SM}(x)$ in (3.1b).

As we have seen, superspace reformulation has many advantages, not only providing a clearer overall picture, but also technical details, such as the closure of gauge algebra which we skipped in the component formulation.

## 8. Concluding Remarks

In this paper, we have established $N=1$ supersymmetric compensator SM for an arbitrary non-Abelian gauge group. We have seen that the field strength $P_\mu{}^I$ plays a crucial role in the supersymmetric couplings. We have also seen the peculiar terms such as those in the covariant derivative $\mathcal{D}_\mu \chi$, in which the $P\chi$-term plays important roles for supersymmetric consistency. We have also seen that we can stay within the Wess-Zumino gauge for the VM in order to handle the compensator couplings. This feature is also important to build simple lagrangians and transformation rules. We have further seen that the $m \to 0$ limit corresponds to the supersymmetric $\sigma$-model on the group manifold for $G$. We have seen how the compensator field strength is related to $\sigma$-model notations, such as the metric of the group manifold.

---

[9] Since we are using anti-hermitian generators in the expansion $X \equiv X^I T^I$ instead of hermitian ones, we have associated differences from [17] in these coefficients.

As the most important application, we have presented non-Abelian self-dual VM, in which the compensator SM solves the conventional problem for non-Abelian couplings. This is because we can keep gauge invariance as the guiding principle to fix the supersymmetric lagrangian. We have seen that the quantization of the $\mu$-coefficient as purely non-Abelian effect, which did not arise in the Abelian case [7]. Interestingly, we see that the mass of the self-dual vector itself is quantized, by the identification $\mu \equiv m$.

As additional by-products, we have also performed the coupling to supersymmetric DBI actions at the first non-trivial order. These generalizations have been supposed to be extremely difficult, since the discovery of the self-dual VM in 3D [7]. These are all extra new results, following our successful formulation of $N = 1$ supersymmetric compensator SM in 3D.

We have further reformulated all the component results in superspace, which provides a more transparent picture of the total subject. We stress that even though the scalar multiplet $(\varphi, \chi; F)$ has been known for decades since 1970's, it is shown for the first time that this multiplet serves as a compensator multiplet for non-Abelian gauge groups. These results both in superspace and components indicate that there are still lots of unknown features in supersymmetry, such as the new off-shell transformation rule (3.3d) with the $\chi^2$-term, even thirty years after its first discovery [17][18].

Our results also show that 3D are very advantageous for dealing with compensator SMs, because a SM in 3D has the simple structure $(\varphi, \chi; F)$ only with one physical scalar field. For example, in 4D we have already mentioned the obstruction for supersymmetric non-Abelian compensator, due to the two different spin 0 fields interfering. In principle, we can try similar procedure for compensator SM in higher dimensions. However, it seems impossible to build a similar theory in 7D or higher, because of the lack of SMs in those dimensions [19]. In this context, the particular importance of 3D should be also emphasized, as the base manifold for supermembrane theory [20].

It has been well known that self-dual Yang-Mills theory in 4D [21], together with its supersymmetric versions [22][23], have close relationships with integrable models in lower dimensions. From this viewpoint, it is natural to expect similar relationships between a supersymmetric 'self-dual' Yang-Mills in 3D and supersymmetric integrable models in lower dimensions. With the non-Abelian self-dual VM established at hand, we are now at a better position to investigate this important question with supersymmetry, self-duality and non-Abelian interactions.

This work is supported in part by NSF Grant # 0308246.